\documentclass[longauth, letter]{aa}  

\usepackage{graphicx}
\usepackage{txfonts}
\usepackage{lipsum}
\usepackage{subcaption}
\usepackage{lscape}
\usepackage{placeins}

\usepackage{hyperref}

\usepackage[LGR, T1]{fontenc}

\DeclareTextAccentDefault{\accdasia}{LGR}
\DeclareTextAccentDefault{\acctonos}{LGR}

\usepackage{xcolor}
\usepackage{enumitem}

\hypersetup{colorlinks=true, citecolor=blue, linkcolor=blue}

\newcommand{\clump}{Hebe}

\newcommand{\heiiuv}{He{\sc{ii}}$_{1640}$}
\newcommand{\heii}{He{\sc{ii}}}

\newcommand{\oiib}{[O{\sc{ii}}]$_{3727}$}
\newcommand{\oiir}{[O{\sc{ii}}]$_{3730}$}

\newcommand{\ha}{H$\alpha$}

\newcommand{\hg}{H$\gamma$}
\newcommand{\hd}{H$\delta$}
\newcommand{\he}{H$\epsilon$}
\newcommand{\oiiia}{[O{\sc{iii}}]$_{4364}$}

\newcommand{\neiiib}{[Ne{\sc{iii}]}$_{3870}$}

\newcommand{\micron}{\ensuremath{\mu\mathrm{m}}\xspace}

\defcitealias{Maiolino26b}{M26}
\defcitealias{Maiolino24b}{M24}
\defcitealias{Nakajima22}{N22}
\defcitealias{Rusta25}{R25}
\defcitealias{Bunker23}{B23}

\begin{document}

   \title{GA-NIFS and JADES: Confirmation of pristine gas near GN-z11}

    \author{H.~{\"U}bler \inst{\ref{MPE}}
   \and 
   R.~Maiolino\inst{\ref{Kavli},\ref{Cavendish},\ref{UCL}} \and
    P.~G.~Pérez-González\inst{\ref{CAB}} \and
    Y.~Isobe\inst{\ref{Kavli},\ref{Cavendish},\ref{Waseda}} \and
    G.~C.~Jones\inst{\ref{Kavli},\ref{Cavendish}} \and
    N.~Kumari\inst{\ref{ESA-Baltimore}} \and
    S.~Charlot\inst{\ref{CNRS}} \and
    E.~Rusta\inst{\ref{Florence},\ref{INAF-F}} \and  
    S.~Salvadori\inst{\ref{Florence},\ref{INAF-F}} \and
    K.~Nakajima\inst{\ref{Kanasawa}, \ref{Kanasawa2}, \ref{NAOJ}} \and
    M.~Perna\inst{\ref{CAB}} \and    
    S.~Arribas\inst{\ref{CAB}} \and
    A.~J.~Bunker\inst{\ref{Oxford}} \and
    S.~Carniani\inst{\ref{SNS}} \and
    F.~D'Eugenio\inst{\ref{Kavli},\ref{Cavendish}} \and
    B.~Rodríguez Del Pino\inst{\ref{CAB}} \and    
    E.~Bertola\inst{\ref{INAF-F}} \and
    T.~B\"oker\inst{\ref{STScI}} \and
    J.~Chevallard\inst{\ref{Oxford}} \and
    C.~Circosta\inst{\ref{IRAM}} \and
    G.~Cresci\inst{\ref{INAF-F}} \and
    M.~Curti\inst{\ref{INAF-OAS}} \and 
    E.~Curtis-Lake\inst{\ref{Herts}} \and
    D.~J.~Eisenstein\inst{\ref{CfA}} \and
    K.~Hainline\inst{\ref{Arizona}} \and
    B.~D.~Johnson\inst{\ref{CfA}} \and
    E.~Parlanti\inst{\ref{SNS}} \and 
    P.~Rinaldi\inst{\ref{STScI}} \and
    B.~Robertson\inst{\ref{UCSC}} \and
    J.~Scholtz\inst{\ref{Kavli},\ref{Cavendish}} \and
    S.~Tacchella\inst{\ref{Kavli},\ref{Cavendish}} \and
    G.~Venturi\inst{\ref{SNS}} \and
    C.~N.~A.~Willmer\inst{\ref{Arizona}} \and
    J.~Witstok\inst{\ref{DAWN},\ref{Copenhagen}} \and
    S.~Zamora\inst{\ref{SNS}}
        } 

   \institute{
   Max-Planck-Institut f\"ur extraterrestrische Physik, Gie{\ss}enbachstra{\ss}e 1, 85748 Garching, Germany; {\tt{hannah@mpe.mpg.de}}\label{MPE}  \and
    Kavli Institute for Cosmology, University of Cambridge, Madingley Road, Cambridge, CB3 0HA, UK\label{Kavli} \and
    Cavendish Laboratory, University of Cambridge, 19 JJ Thomson Avenue, Cambridge, CB3 0HE, UK\label{Cavendish} \and   
    Department of Physics and Astronomy, University College London, Gower Street, London WC1E 6BT, UK\label{UCL} \and
    Centro de Astrobiolog\'ia (CAB), CSIC--INTA, Cra. de Ajalvir km.~4, 28850 -- Torrej\'on de Ardoz, Madrid, Spain\label{CAB} \and
    Waseda Research Institute for Science and Engineering, Faculty of Science and Engineering, Waseda University, 3-4-1, Okubo, Shinjuku, Tokyo 169-8555, Japan\label{Waseda} \and
    AURA for European Space Agency (ESA), ESA Office, Space Telescope Science Institute, 3700 San Matin Drive, Baltimore, MD, 21218, USA\label{ESA-Baltimore} \and
    Sorbonne Universit\'e, CNRS, UMR 7095, Institut d'Astrophysique de Paris, 98 bis bd Arago, 75014 Paris, France\label{CNRS} \and
    Dipartimento di Fisica e Astronomia, Università degli Studi di Firenze, Largo E. Fermi 1, 50125, Firenze, Italy\label{Florence} \and
    INAF — Osservatorio Astrofisico di Arcetri, Largo E. Fermi 5, I-50125, Florence, Italy\label{INAF-F} \and
    Institute of Liberal Arts and Science, Kanazawa University, Kakuma-machi, Kanazawa, 920-1192, Ishikawa, Japan\label{Kanasawa} \and
    Division of Mathematical and Physical Sciences, Graduate School of Natural Science and Technology, Kanazawa University, Kakuma-machi, Kanazawa, 920-1192, Ishikawa, Japan\label{Kanasawa2} \and
    National Astronomical Observatory of Japan, 2-21-1 Osawa, Mitaka, 181-8588, Tokyo, Japan\label{NAOJ} \and
    Department of Physics, University of Oxford, Denys Wilkinson Building, Keble Road, Oxford OX1 3RH, UK\label{Oxford} \and
    Scuola Normale Superiore, Piazza dei Cavalieri 7, I-56126 Pisa, Italy\label{SNS} \and
    European Space Agency, c/o STScI, 3700 San Martin Drive, Baltimore MD 21218, USA\label{STScI} \and
    Institut de Radioastronomie Millim\'etrique (IRAM), 300 Rue de la Piscine, 38400 Saint-Martin-d’H\`eres, France\label{IRAM} \and
    INAF - Osservatorio di Astrofisica e Scienza dello Spazio, Via Piero Gobetti 93/3, 40129, Bologna, Italy\label{INAF-OAS} \and
    Centre for Astrophysics Research, Department of Physics, Astronomy and Mathematics, University of Hertfordshire, Hatfield AL10 9AB, UK\label{Herts} \and 
    Center for Astrophysics $|$ Harvard \& Smithsonian, 60 Garden St., Cambridge MA 02138 USA\label{CfA} \and
    Steward Observatory, University of Arizona, 933 N. Cherry Avenue, Tucson, AZ 85721, USA\label{Arizona} \and
    Department of Astronomy and Astrophysics, University of California, Santa Cruz, 1156 High Street, Santa Cruz, CA 95064, USA\label{UCSC} \and
    Cosmic Dawn Center (DAWN), Copenhagen, Denmark\label{DAWN} \and
    Niels Bohr Institute, University of Copenhagen, Jagtvej 128, DK-2200, Copenhagen, Denmark\label{Copenhagen}
            }

  \abstract 
  {
  According to the leading cosmological model, a first generation of stars called Population~III (PopIII), condensed almost entirely out of hydrogen and helium, must have initiated the creation of all heavier chemical elements. 
  We report the detection of ionised hydrogen (H$\gamma_{4342}$) with a signal-to-noise ratio of $S/N$=5.9 in a region about 3~pkpc (projected) north-east from the $z$$\sim$10.6 galaxy GN-z11, where line emission compatible with doubly ionised helium (\heiiuv) has been found. 
  Our new JWST/NIRSpec-IFU G395H data confirm the authenticity of the previous detection at a redshift of $z_{\rm H\gamma}$=$10.5862$$\pm$$0.0003$. 
  \hd\ is marginally detected ($S/N$$\sim$$2$). 
  No metal lines are detected in our observations spanning $\lambda_{\rm rest}$=$0.25$-$0.45\mu$m.
  We derive a $3\sigma$ upper limit on the gas phase metallicity of 12+log(O/H)$<$7.0 ($Z_{\rm gas}$$<$$0.02~Z_\odot$).
  Through comparison with NIRCam imaging, we constrain a lower limit on the equivalent width of EW$_0$(\hg)$>$350\AA.
  We compare our emission line constraints to model predictions and find them compatible with photoionisation by PopIII stars, possibly intermixed with next-generation (PopII) stars.
  We infer an upper limit on the dynamical mass of $M_{\rm dyn}$$\lesssim$$3$$\times$$10^8M_\odot$.
  Our data provide novel support for the presence of PopIII stars nearby GN-z11, 440~Myr after the Big Bang.
  }

   \keywords{galaxies: high-redshift -- stars: Population III -- galaxies: formation}

   \maketitle

\nolinenumbers

\section{Introduction}

A key prediction of the leading cosmological model is the formation of the first chemical elements during Big Bang nucleosynthesis: hydrogen, helium, and only trace amounts of few heavier elements \citep{Pitrou18}. All further production of metals hinges on the formation of the very first generation of stars (PopIII) that initiate and promote the production of elements heavier than helium through stellar nucleosynthesis and stellar explosions \citep{Johnson19}.
It is a decisive observational goal to test and constrain current theories of the early evolution of our Universe by finding and characterising this first generation of stars.
The detection of the signatures of these first stars through their ionising effect on surrounding pristine gas is impeded by their potentially short lifetimes \citep{Bromm04}. Finding a population of first stars mixed with next-generation (PopII) stars may be more likely \citep[e.g.][hereafter R25]{Rusta25}.

After almost four years of science operations of the {\it James Webb} Space Telescope (JWST), a handful of such candidate systems have been identified at 4$<$$z$$<$7 based on photometry or on the detection of hydrogen lines with no or weak accompanying metal lines. These systems that are characterised by low gas-phase metallicities \citep[12+log(O/H)$<$6.5; $Z$$\lesssim$$10^{-2}Z_\odot$;][]{Vanzella23, Vanzella26, Morishita25, Maiolino25b, Fujimoto25}. However, even a non-detection of metal lines may be insufficient to claim a PopIII population \citep{Katz23, Storck25}, for which $Z$$\lesssim$$10^{-5}$-$10^{-6}Z_\odot$ is expected \citep{Klessen23}. A more accessible probe is doubly ionised helium. In combination with ionised hydrogen and absent metal lines, it can place constraints on the hardness of the incident radiation field \citep{Tumlinson00, Schaerer02, Nakajima22, Katz23, Lecroq25}.
In this context, the tentative detection of \heiiuv\ from medium-resolution ($R$$\sim$$1000$) data \citep[][hereafter M24]{Maiolino24b} close to the $z$=$10.6$ galaxy GN-z11 (\citealp{Oesch16}; \citealp{Bunker23}, hereafter B23; \citealp{Tacchella23, Maiolino23a, AlvarezMarquez25}) is quite intriguing. Because no accompanying emission lines were found, the nature and redshift of the line emitter remained ambiguous. 

In this Letter, we present new high-resolution (G395H; $R$$\sim$2700) Near Infrared Spectrograph Integral Field Unit \citep[NIRSpec-IFU;][]{Jakobsen22, Boeker22} data of this system as part of the GTO survey \href{https://ga-nifs.github.io/}{`Galaxy Assembly with NIRSpec-IFS'} (GA-NIFS) under program 4528 (PIs: S.~Arribas, R.~Maiolino, K. Isaak). We report the detection of \hg\ and a marginal detection of \hd\ at the putative \heii\ clump location, without any metal lines. Our \hg\ detection confirms the authenticity of the previous \heiiuv\ detection. A companion paper by \citeauthor{Maiolino26b} (2026, hereafter M26) corroborates the \heiiuv\ data through deeper NIRSpec-IFU follow-up.
We derive new constraints on the metallicity and nature of the line emitter, which we dubbed `\clump'.

\section{Observations and data processing}\label{s:data}

\begin{figure}
\centering
        \includegraphics[width=0.9\columnwidth]{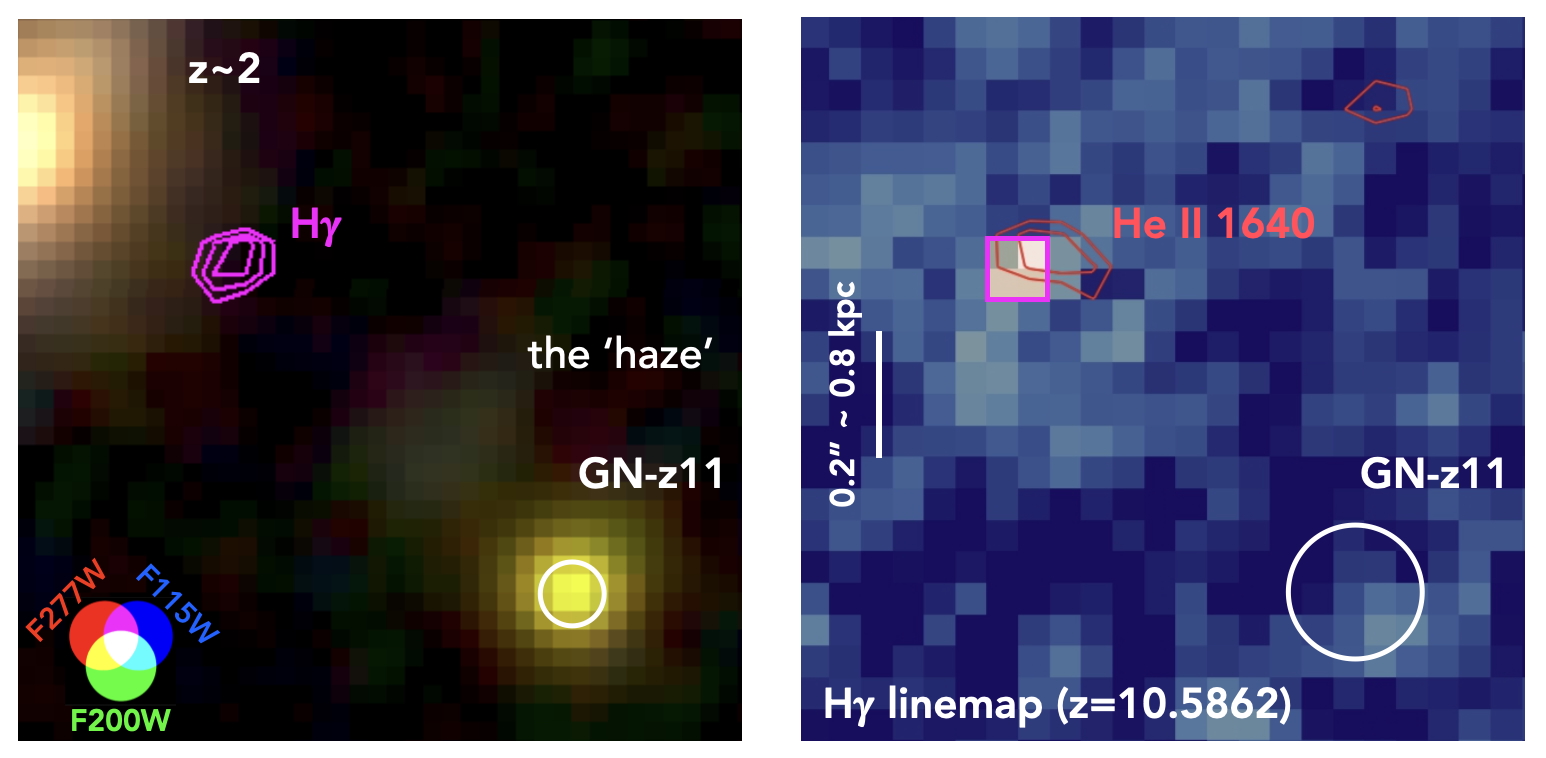}
    \caption{\small {\it Left:} Near Infrared Camera (NIRCam) {\it rgb} image showing the location of the \hg\ clump (pink contours at $3\sigma$, $4\sigma$, $5\sigma$) in reference to GN-z11, the haze \citep{Tacchella23}, and a $z$$\sim$$2$ galaxy. The circle indicates the astrometric uncertainty ($r$=$0.05^{\prime\prime}$) in registering the NIRCam images and NIRSpec-IFU cubes, comparable to the NIRCam PSF at 2.8\micron\ ($\sim$0.09\arcsec).
    {\it Right:} \hg\ line map ($\lambda_{\rm obs}$=5.029-5.031\micron). The pink square shows the extraction aperture for Fig.~\ref{f:fit_hg}. The red contours (at $4\sigma$, $5\sigma$) trace \heiiuv\ by \citetalias{Maiolino26b}. The circle shows the NIRSpec PSF at 5.3\micron.
    }
    \label{f:map_extraction}
\end{figure}

Our new NIRSpec-IFU data cover GN-z11 and its immediate environment within $\sim$6~pkpc. The data were taken on April 9, 2025, with an eight-dither medium cycling pattern and a total integration time of 6.9~h with the grating/filter pair G395H/F290LP, covering the wavelength range $2.87$-$5.27\mu$m.
Raw data were processed with the JWST Science Calibration pipeline version 1.15.0 and jwst\_1293.pmap. To increase the data quality, we performed additional reduction steps described in detail by \cite{Perna23} and summarised here. 
Count-rate frames were corrected for $1/f$ noise through a polynomial fit. 
We manually removed regions affected by failed-open MSA shutters and cosmic-ray residuals.
The remaining outliers were rejected with a 95th percentile threshold following the procedure by  \cite{DEugenio23}.
The final cube was combined using the drizzle method with a pixel scale of $0.05''$.
We used spaxels that were free of emission to perform a background subtraction. 
As noted in previous works, the {\sc{err}} extension in the combined cube underestimates the noise. Following \cite{Uebler23}, we rescaled the formal uncertainties to match the rms in regions free of spectral features, here, by a factor $2.2$.
To analyse the line emission, we also performed a local continuum subtraction to account for any contamination to the continuum by a foreground $z$$\sim$2 galaxy (Fig.~\ref{f:map_extraction}, left; see \citealp{Tacchella23}; see App.~\ref{a:contsub} for details). No line emission is seen or expected at the wavelength of our $z$$\sim$$10.6$ \hg\ detection from this low$-z$ galaxy. We accounted for lensing magnification by the foreground galaxy of $\mu$$\sim$$1.42$ throughout the paper \citepalias[see][for details]{Maiolino26b}.

\section{Results}

\subsection{Confirmation of a metal-poor line emitter at \texorpdfstring{$z=10.5862$}{z=10.5862}}\label{s:metal}

\begin{figure*}
\centering
        \includegraphics[width=0.66\textwidth]{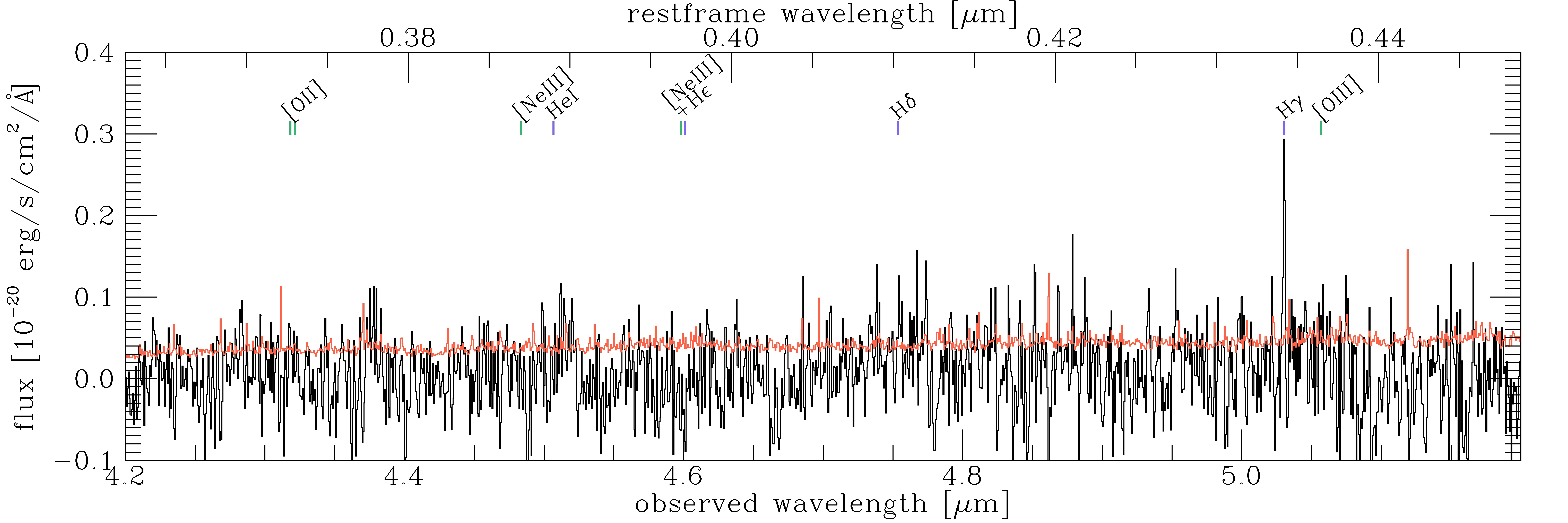}
    \includegraphics[width=0.31\textwidth]{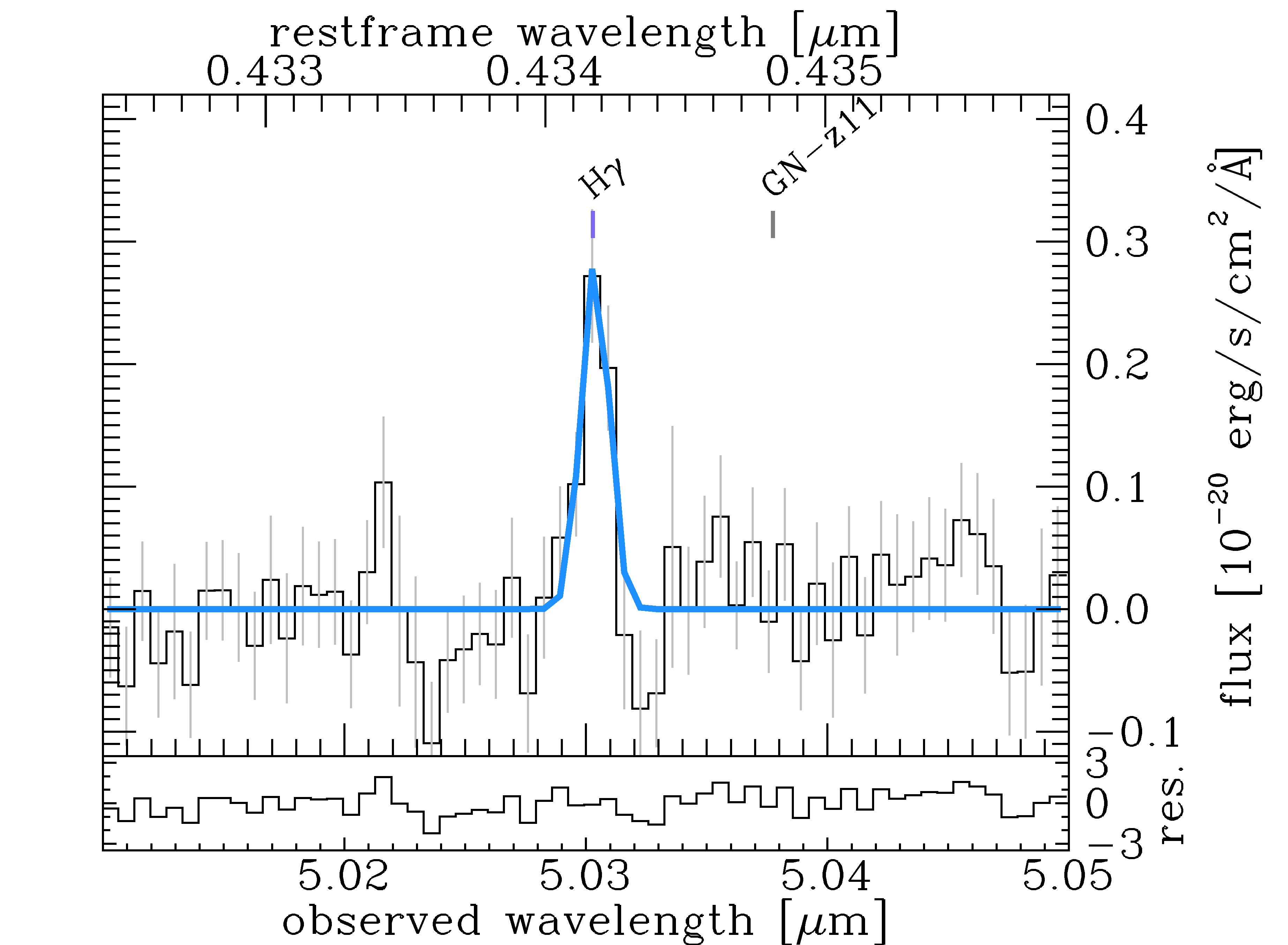}
    \caption{\small 
    {\it Left:} Integrated spectrum of the \hg\ clump (pink aperture in Fig.~\ref{f:map_extraction}). The error spectrum is shown in red. We indicate the theoretical locations of several emission lines at $z$$=$$10.586$.
    {\it Right:} Best fit to \hg, detected with $S/N$$\sim$$6$. In grey we mark the wavelength of \hg\ in GN-z11.
    }
    \label{f:fit_hg}
\end{figure*}

We find an emission line at $\lambda_{\rm obs}$=$5.03\mu$m with high significance ($S/N$=5.9) at the location of the previous putative \heiiuv\ detection by \citetalias{Maiolino24b}, 3~kpc (projected) north-east of GN-z11 (Fig.~\ref{f:map_extraction}). This emission matches \hg\ at $z$=$10.5862$$\pm$$0.0003$.
The right panel of Fig.~\ref{f:map_extraction} shows the \hg\ line map collapsed over $\lambda_{\rm obs}$=$5.029$-$5.031\mu$m.
The red contours trace the \heiiuv\ emission derived from the G235H IFU data presented by \citetalias{Maiolino26b}.
The emission peaks of \hg\ and \heii\ are within less than a pixel from each other, which is consistent  within the relative positional uncertainties ($\sim$$0.05^{\prime\prime}$) related to astrometric registration and relative shifts in the pixel grids of the IFU cubes. 
Our data thus provide the decisive second line detection to confirm the hitherto ambiguous He{\sc{ii}} line.
The left panel of Fig.~\ref{f:fit_hg} shows the integrated spectrum extracted from the $0.1^{\prime\prime}$$\times$$0.1^{\prime\prime}$ aperture marked in Fig.~\ref{f:map_extraction}, encompassing the four brightest \hg\ pixels.\footnote{
We note tentative emission at slightly different velocities south of the \hg-emitter (not shown) that requires deeper data for confirmation.
}
We indicate the theoretical $z$=$10.586$ locations of emission lines detected in GN-z11 \citepalias{Bunker23}. \hg\ is clearly detected. The right panel shows a one-component Gaussian fit to \hg\ with local continuum subtraction. The line fluxes and upper limits are reported in Tab.~\ref{t:fluxes}.

\begin{table}
\tiny
\caption{\small $R$$\sim$$2700$ emission line fluxes and upper limits extracted from the integrated spectrum (Fig.~\ref{f:fit_hg}). 
}\label{t:fluxes}
\begin{tabular}{@{}lcc@{}}
Line & Flux ($0.1^{\prime\prime}$$\times$$0.1^{\prime\prime}$)  & Flux (magn. \& ap. corr.) \\
&   [$10^{-20}$~erg/s/cm$^2$] & [$10^{-20}$~erg/s/cm$^2$] \\
\hline
\hg & $4.1\pm0.7$ & $16.0\pm2.7$ \\ 
\hd & $1.2\pm0.6$ & $4.4\pm2.2$ \\ 
\he & $<3.1$ & $<11.5$ \\ 
He~{\sc{i}}$_{3890}$ & $<2.0$ & $<7.0$ \\ 
\oiiia & $<3.8$ & $<14.9$ \\ 
\neiiib & $<1.8$ & $<6.3$ \\ 
\oiir & $<2.1$ & $<7.4$ \\ 
\oiib & $<2.5$ & $<8.8$ \\ 
\hline
EW$_0$(\hg) [\AA ] & -- & $>350$ \\
\hline
\end{tabular}
\tablefoot{ 
The flux uncertainties were calculated from the local noise. For non-detections, we calculated $3\sigma$ upper limits from 500 Monte Carlo realisations of the spectrum. The third column provides the fluxes corrected for magnification ($\mu$$\sim$$1.42$) by the foreground galaxy and for aperture losses assuming a point-source geometry for the line-emitting region. The lower limit for EW$_0$(\hg) was calculated by comparing the aperture-corrected flux to the 5$\sigma$ upper limit continuum magnitude.}
\end{table}

\hg\ is blueshifted with respect to the systemic velocity of \hg\ in GN-z11 by about $-390$~km/s.\footnote{In App.~\ref{a:bound} we explore whether Hebe is bound to GN-z11.} 
\citetalias{Maiolino24b} found $z_{\rm HeII}$=$10.600$, corresponding to a shift of about $-80$~km/s, based on lower-$S/N$ $R$$\sim$$1000$ data. The deeper G235H data reported by \citetalias{Maiolino26b} show a more complex double-peaked \heii\ spectral profile, where the red peak (component C2) is consistent with our \hg\ redshift and emission (see Fig.~\ref{f:hgheiivel} and \citetalias{Maiolino26b}).

The detection of \hg\ from \clump\ unambiguously validates the line detection by \citetalias{Maiolino24b} as \heiiuv.
Our measurement of the total \hg\ flux ($16.0$$\pm$$2.7$$\times$$10^{-20}$erg/s/cm$^2$) is higher than the $3\sigma$ upper limit derived from an offset NIRSpec-MSA aperture (<$1.37$$\times$$10^{-19}$erg/s/cm$^2$) by \citetalias{Maiolino24b}; this is expected since the MSA aperture covers only part of the \heii+\hg\-emitting region \citepalias[see][]{Maiolino24b}. 
\hd\ is only marginally detected in our data ($S/N$$\sim$$2$), and we find \hg/\hd=3.6$\pm$$1.9$. This is higher than but still compatible with the uncertainties with the expectation for dust-free case B (\hg/\hd=1.8 for $T_e$=$10^4$~K, $n_e$=$10^4{\rm cm}^{-3}$). 
If \hd\ is indeed weaker than expected, several reasons might explain this: lower temperatures or higher densities might increase \hg/\hd. Extremely large Balmer optical depths \citep[case C recombination; e.g.][]{Xu92} might alter \hg/\hd, although such conditions generally require (column) densities higher than expected in H\,{\sc ii} regions. Another reason might be  dust, which can exist even in PopIII clusters through supernova self-enrichment prior to the condensation of PopII stars \citepalias[e.g.][]{Rusta25}. 

Our data cover the theoretical locations of metal lines from Mg{\sc{ii}}$_{2795,2802}$ to \oiiia. In GN-z11, the brightest of these is \neiiib\ \citepalias{Bunker23}. Fig.~\ref{f:fit_hg} shows that these lines are not detected at the location of \clump. 
We used the (magnification- and aperture-corrected) $3\sigma$ upper limit on \neiiib\ to derive an upper limit on the gas-phase metallicity: using the high$-z$ calibration by \cite{Isobe26}\footnote{
The authors derive the relation between temperature-based 12+log(O/H) and \neiiib/\hg\  from stacks of $z$=1-10 spectra from JADES \citep{Scholtz25b} and Dark Horse \citep{DEugenio25c} in bins of 12+log(O/H) based on the strong-line method \citep{Cataldi25} and stellar mass.}
for \neiiib/\hg\, we found 12+log(O/H)$<$7.0, i.e.~$Z_{\rm gas}$$<$$0.02Z_\odot$.
Our \hg\ detection and confirmation of this metal-poor clump adds to the growing literature of metal-poor or pristine gas conditions during the first few billion years  \citep{Vanzella23, Vanzella26, Cullen25, Morishita25, Maiolino25b, Willott25, AlvarezMarquez26, Reumert26}.

We calculated the \hg\ equivalent width using our \hg\ flux and a continuum measurement extracted from NIRCam imaging. We first measured the flux in the F444W filter in the pink aperture (Fig.~\ref{f:map_extraction}) and corrected it for a point-source geometry, finding $3.6$$\pm$$0.8$~nJy. To avoid contamination from the outskirts of the nearby $z$$\sim$2 galaxy, we subtracted an isophotal model of that galaxy from the F444W image, repeated the measurement, and obtained $1.0$$\pm$$0.8$~nJy. Given the low $S/N$ of this measurement, we hereafter consider the corresponding 5$\sigma$ upper limit by deriving the noise in a $5$$\arcsec$$\times$$5$$\arcsec$ region around \clump, yielding 3.9~nJy. We accounted for contribution of \hd\ and \he\ to the broadband flux (using the \he\ upper limit and assuming case B) and adopted a flat spectrum between F444W and $\lambda_{\rm obs,H\gamma}$, resulting in a 5$\sigma$ upper limit for the magnification-corrected spectral continuum rest-frame flux of $4.6$$\times$$10^{-22}$erg/s/cm$^2$/\AA. This implies a lower limit on the rest-frame \hg\ equivalent width of EW$_0$(\hg)$>$350\AA.

\begin{figure*}[t]
\centering
        \includegraphics[width=0.91\columnwidth]{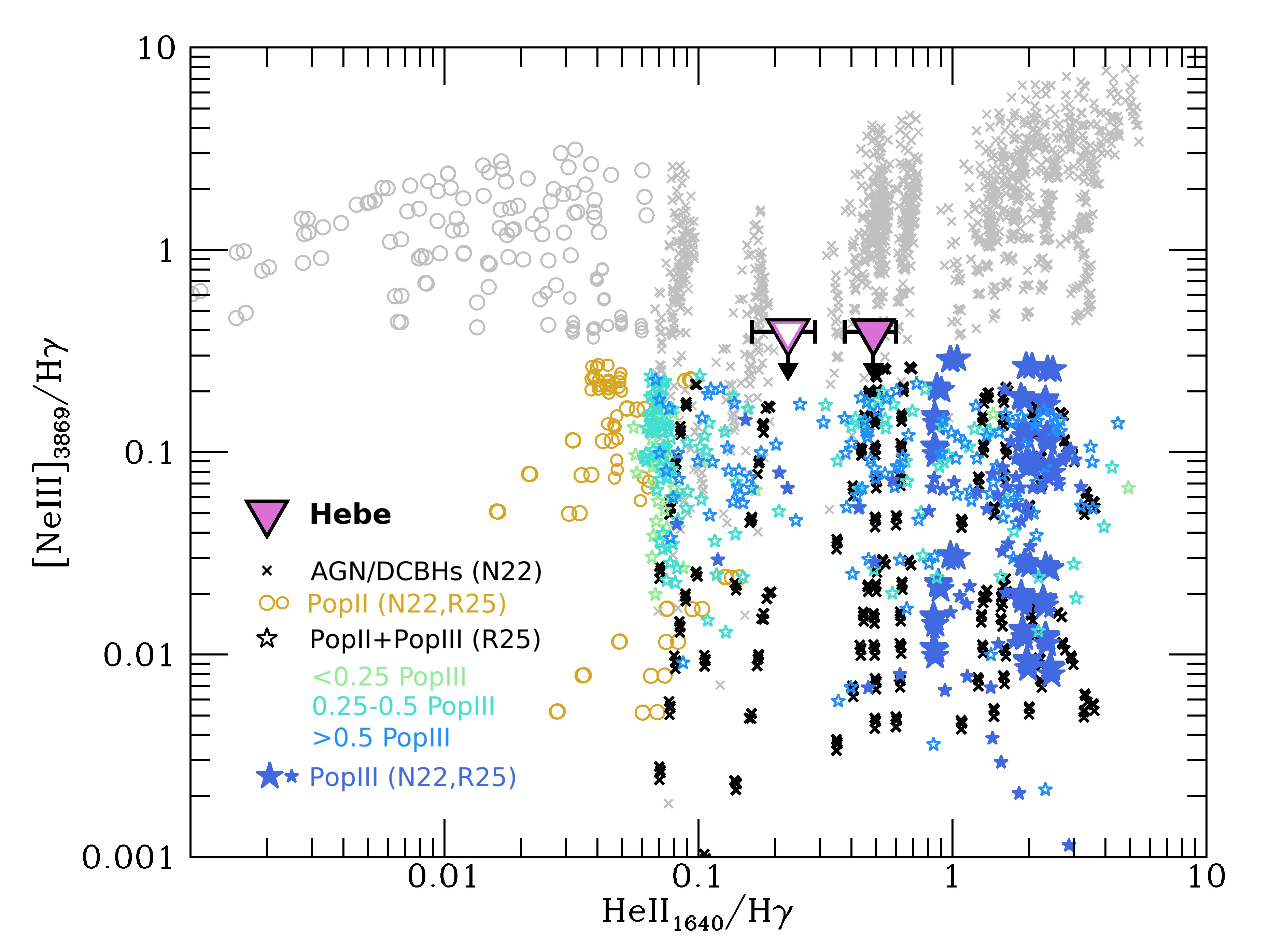}
        \includegraphics[width=0.91\columnwidth]{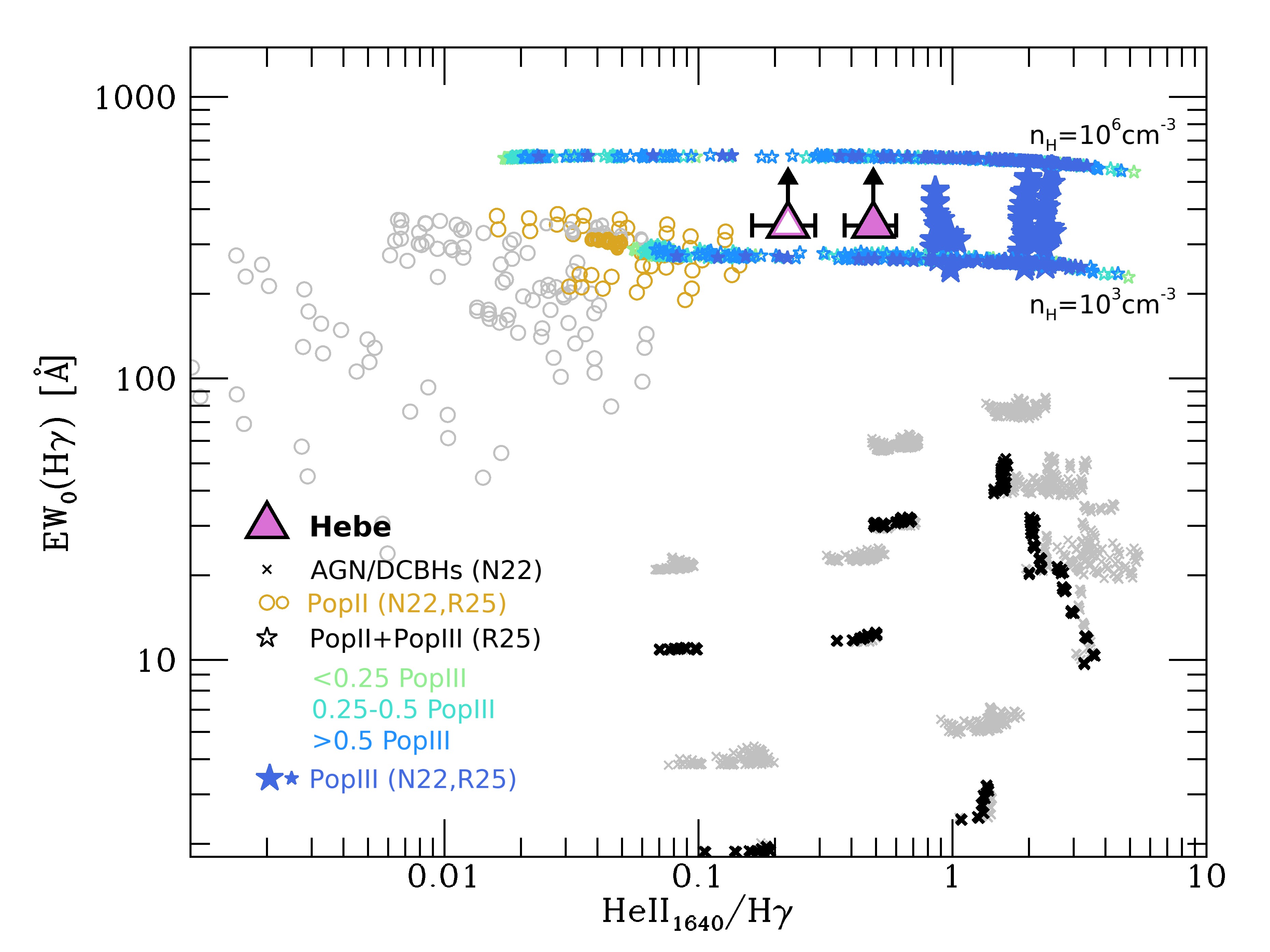}
    \caption{\small Diagnostic diagrams of \neiiib/\hg\ ({\it left}) and EW$_0$(\hg) ({\it right}) vs. \heiiuv/\hg. The pink triangles show our constraints for \clump\ from this work and \citetalias{Maiolino26b} using either the total \heii\ flux (filled) or the \heii\ component matching our \hg\ detection (open; see \citetalias{Maiolino26b} and App.~\ref{a:velmatch}). 
    The other large and small symbols show model predictions by \citetalias{Nakajima22} and \citetalias{Rusta25}, respectively, for ionisation by PopIII (filled stars), mixed PopIII+PopII (open stars; fractions of total stellar mass), PopII (open circles), and AGN or DCBHs (crosses), as indicated in the legend.
    For PopIII stars, the models shown have stellar masses $M_\star$=$1$-$500 M_\odot$, absolute gas-phase metallicity $Z$=0-$10^{-4}$, neutral hydrogen density $n_H$=$10^{3-6}$cm$^{-3}$, ionisation parameter $\log(U)$=$[-0.5;-2]$ for \citetalias{Nakajima22}, and $M_\star$=$0.8$-$1000 M_\odot$, $Z$=0-$2.7$$\times$$10^{-4}$, $n_H$=$10^3$cm$^{-3}$, $\log(U)$=$-1$ (and $n_H$=$10^6$cm$^{-3}$, $\log(U)$=$-2$ in the right panel) for \citetalias{Rusta25}. For AGN/DCBH, we show models with $n_H$=$10^{3-6}$cm$^{-3}$.
    Models with gas metallicities that fail to meet our upper limit are greyed out.
    }
    \label{f:diag}
\end{figure*}

We note that any \ha\ emission from \clump\ is covered in the MIRI-MRS observations presented by \cite{AlvarezMarquez25}. For case~B conditions and no dust, the noise level of the MIRI data matches the \ha\ flux expected from our measurement of \hg. Using the $1\sigma$ flux uncertainty (derived for GN-z11 by \citealp{AlvarezMarquez25}) and a point-source geometry for \clump, we estimate that an 11$\times$ longer MIRI exposure would be required to detect \ha\ at the clump position.

\subsection{The case for PopIII stars}

The detection of hydrogen and helium lines without emission from heavier chemical elements could be a signpost for the long-sought first generation of stars. In Fig.~\ref{f:diag} we place the new constraints from our data and \citetalias{Maiolino26b} in context with model predictions by \citeauthor{Nakajima22} (2022; hereafter N22) and \citetalias{Rusta25}. On the left, we plot \heiiuv/\hg, a measure for the hardness of the ionising spectrum, versus \neiiib/\hg, essentially constraining the metallicity. The \heii\ fluxes are taken from \citetalias{Maiolino26b} (filled triangle for the total flux, open triangle for their component C2 that matches our \hg\ detection; see App.~\ref{a:velmatch} and \citetalias{Maiolino26b}), while the \hg\ and \neiiib\ constraints come from this work. 
On the right, we show EW$_0$(\hg) versus \neiiib/\hg.
If we accounted for the effects of dust, the predictions would be moved towards lower \heiiuv/\hg\ \citep[see][]{Rusta26}.
Ionisation by evolved (PopII) stars is ruled out through the high \heiiuv/\hg, based on the models we considered. 
Instead, our data are compatible with ionisation by (self-polluted) PopIII stars or by a hybrid PopIII+PopII system \citepalias{Rusta25}, potentially with high gas densities.
The comparison of our data and the model predictions highlights the potential role of mixed PopIII+PopII systems in producing the metal-poor emission discussed here and in the recent literature \citep{Vanzella23, Vanzella26, Morishita25}.

In the \heiiuv/\hg\ versus \neiiib/\hg\ parameter space, some AGN/direct collapse black hole \citep[DCBH; e.g.][]{Loeb94, Bromm03} models by \citetalias{Nakajima22} are also compatible with our data.
However, the high EW$_0$(\hg) effectively rules out a dominant contribution by AGN (right panel), even more strongly than the EW$_0$(\heii) discussed by \citetalias{Maiolino26b}.
A black hole embedded in a star cluster might still contribute to the ionisation.
We found no evidence for a broad \hg\ component in Hebe, which might indicate the broad-line region of an AGN, underneath the narrow component. However, the narrow component itself could trace the broad-line emission by a low-mass AGN. 
The \hg\ line width is comparable to the instrumental spectral resolution at that wavelength \citep{Jakobsen22}. Thus, we used the nominal instrumental resolution of $\sigma_{\rm inst.,H\gamma}$$\sim$$37$~km/s as an upper limit on the velocity dispersion. 
Assuming case B conditions and local scaling relations to convert line luminosity and width to black hole mass \citep{Greene05, Reines15}, we calculated an upper limit of the putative black hole mass in the range $M_\bullet$$<$$(0.6$-$10)$$\times$$10^4M_\odot$. This estimate is at the lower end of masses typically associated with DCBHs \citep[see][]{Inayoshi20}.

\subsection{Dynamical mass estimate}

We used the virial theorem to estimate an upper limit on the dynamical mass of the clump: $M_{\rm dyn}$=$C\sigma^2 R/G$, where $C$ is a prefactor depending on the geometry of the system, $\sigma$ is the integrated stellar velocity dispersion, $R$ is the characteristic radius, and $G$ is the gravitational constant. We lack a measurement of the stellar velocity dispersion and used the upper limit on the \hg\ dispersion discussed above to calculate an upper limit on the dynamical mass. 
As the \hg\ emission of \clump\ is unresolved (Fig.~\ref{f:map_extraction}), we used the PSF FWHM as an upper limit on the size. With $R$=${\rm FWHM_{PSF,H\gamma}}$$\sim$$0.21^{\prime\prime}$ \citep{Jones26} and $C$=$1$, we found $M_{\rm dyn}$$<$$2.6$$\times$$10^8 M_\odot$.
Alternatively, we considered that the emission stemmed from a star cluster. The typical effective radii of star clusters are smaller than one to few dozen parsec \citep[e.g.][]{Brown21, Vanzella23b, Adamo24}. We assumed $R$=$10$~pc. For $C$, we adopted 9.75, which is a typical value for globular clusters in the Milky Way \citep[see][and discussion therein]{Bastian06}. With these assumptions, we found $M_{\rm dyn, cluster}$$<$$3.0$$\times$$10^7 M_\odot$.
We note that using the integrated stellar velocity dispersion might increase our estimate of the dynamical mass \citep[see][]{Bezanson18b}.

While uncertain, these upper limits on $M_{\rm dyn}$ are compatible with typical stellar mass estimates for PopIII clusters ($M_{\rm PopIII, cluster}$$\sim$$10^{4-6}M_\odot$; e.g.~\citealp{Venditti23, Storck25, Rusta26}). 
Our upper limit is also compatible with model predictions for the minimum halo mass required for PopIII star formation at $z$$\sim$$10.6$ \citep[see][]{Klessen23}.

\section{Summary and conclusions}

We have presented new JWST/NIRSpec-IFU G395H data on the putative \heiiuv\ clump at a projected distance of about 3~kpc north-east of GN-z11 at $z$$\sim$$10.6$. The goal of this work was to confirm or refute the \heiiuv\ emission and the associated tentative claim of first star signatures by extending the wavelength coverage to $\lambda_{\rm obs}$=$2.87$-$5.27\mu$m, which nominally includes emission lines from Mg{\sc{ii}}$_{2795,2802}$ to \oiiia.
Our results are listed below.

\begin{itemize}[noitemsep,topsep=2pt]

    \item[$\bullet$] We detected \hg\ with an $S/N$$\sim$$6$ at the location of \clump with a redshift of $z_{\rm H\gamma}$=$10.5862$$\pm$$0.0003$. Through this second line detection, we unambiguously confirm the previously reported emission as \heiiuv. 
    
    \item[$\bullet$]Comparing our line flux to a $5\sigma$ upper limit on the continuum, we found a lower limit of EW$_0$(\hg)$>$350\AA.

    \item[$\bullet$] \hd\ is marginally detected ($S/N$$\sim$$2$). Our derived flux is lower than in case B, but consistent within the uncertainties.

    \item[$\bullet$] No heavier chemical elements are detected. Through a $3\sigma$ upper limit on \neiiib\, we derived an upper limit on the gas-phase metallicity of 12+log(O/H)$<$7.0, or $Z_{\rm gas}$$<$$0.02Z_\odot$.

    \item[$\bullet$] Our data are compatible with models predicting ionisation by (self-polluted) PopIII or a mixture of PopIII and PopII stars.

    \item[$\bullet$] We derived a conservative upper limit on the total dynamical mass of the \hg\ clump of $M_{\rm dyn}$$<$$3$$\times$$10^8M_\odot$. 

\end{itemize}
Our new observations strengthen the case for a PopIII clump or a PopIII+PopII system near GN-z11. At $z$=10.5862, \clump\ represents the most distant indication of PopIII stars and pristine or nearly pristine gas found so far. Deeper G395H data would help us to constrain tentative \hg\ emission near \clump\ and enable more detailed studies of the physical properties of this system, including more stringent upper limits on its metallicity.

\bibliographystyle{aa}
\bibliography{literature}

\begin{appendix}

\nolinenumbers

\section{Local continuum subtraction and flux uncertainties}\label{a:contsub}

To analyse line emission we perform a local continuum subtraction, to account for any possible contamination to the continuum by the $z$$\sim$$2$ foreground galaxy (see Section~\ref{s:data}). Specifically, to measure the \hg\ flux we include the local continuum as a constant and fit a Gaussian model plus constant continuum within the wavelength range $\lambda_{\rm obs}$$=$$[5.01,5.05]\mu$m. The flux uncertainty is independently measured from the local noise: we determine the average local noise from the standard deviation of the spectrum in the wavelength range $\lambda_{\rm obs}$$=$$[5.0,5.0286]\mu$m and $[5.032,5.06]\mu$m and multiply by the spectral channel width times the square root of the number of channels contributing to the \hg\ flux as determined from our best fit. 
\hd\ is only marginally detected and we first determine the local continuum in the wavelength range $\lambda_{\rm obs}$$=$$[4.65,4.745]\mu$m and $[4.76,4.85]\mu$m and subtract it before we perform a Gaussian fit where we adopt the line width and the line centroid from our best fit to \hg. We determine the flux uncertainty as for \hg, using the wavelength range $\lambda_{\rm obs}$$=$$[4.65,4.745]\mu$m and $[4.76,4.85]\mu$m to measure the average local noise.

\begin{figure}
\centering
        \includegraphics[width=\columnwidth]{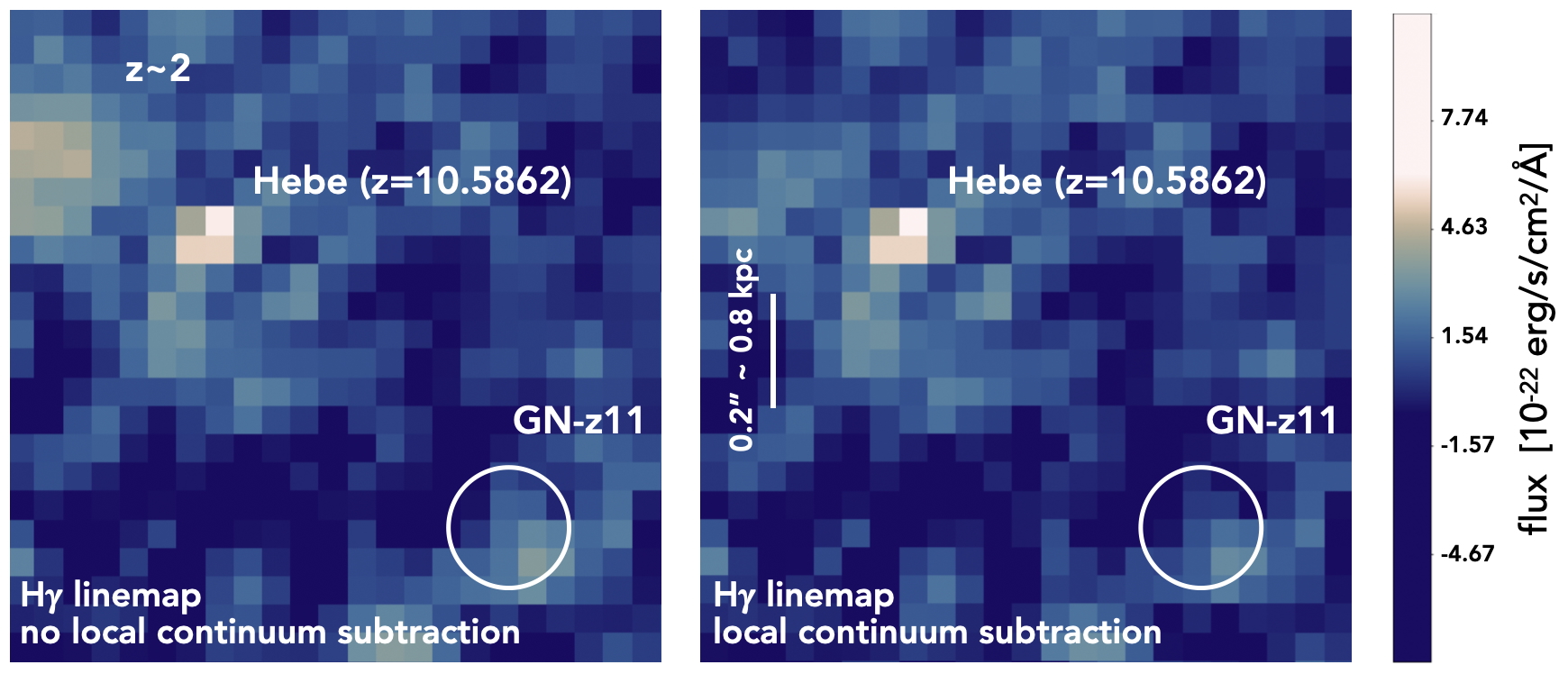}
    \caption{\small \hg\ line map ($\lambda_{\rm obs}$=5.029-5.031\micron) before ({\it left}) and after ({\it right}) local continuum subtraction. The right panel is as in Fig.~\ref{f:map_extraction}. The circle indicating the position of GN-z11 shows the NIRSpec PSF at 5.3\micron. The local continuum subtraction primarily removes possible contamination from the $z$$\sim$2 foreground galaxy.}
    \label{f:comp_cont}
\end{figure}

To obtain the continuum-subtracted line map of \hg\ shown in the right panel of Fig.~\ref{f:map_extraction}, the local continuum is determined in the wavelength range $\lambda_{\rm obs}$$=$$[4.8,5.0]\mu$m and $[5.08,5.24]\mu$m and subtracted, before we collapse the cube in the wavelength range $\lambda_{\rm obs}=5.029-5.031\mu$m. In Fig.~\ref{f:comp_cont} we show the \hg\ line map before (left) and after (right) this local continuum subtraction. This comparison shows that the main effect of the local continuum subtraction is the removal of emission from the $z$$\sim$2 foreground galaxy.

\section{Is Hebe bound to GN-z11?}\label{a:bound}

Hebe is located at a projected distance from GN-z11 of $d$$=$3kpc, with a relative velocity difference of $\Delta v_{\rm los}$$=$$-390$km/s. Understanding whether Hebe is bound to GN-z11 or not could provide important information on the formation pathways of PopIII clusters, specifically whether they form in isolation or whether they can form in pristine gas clumps in the haloes of more massive, evolved galaxies \citep[e.g.][]{Bromm04, Klessen23, Venditti23}. However, because the absolute distance and relative velocity are unknown, as well as the halo mass and profile of GN-z11, we can only attempt an approximate calculation. We assume a halo mass of $M_h=8\times10^{10}M_\odot$ following \cite{Tacchella23} with an NFW profile \citep{NFW96} and a concentration parameter $c$=$3$ \citep[e.g.][]{Correa15}. At 3~kpc, we find an escape velocity of $v_{\rm esc}(r)=\sqrt{2\vert\Phi(r)\rvert}\sim460$~km/s. Thus, if the projected distance and the relative l.o.s.\ velocity are close to the absolute distance and relative velocity, Hebe would be bound to GN-z11. However, if any of the two quantities is larger, which is almost certainly the case, Hebe could easily be unbound (specifically, for a fixed relative velocity $v_{\rm rel}=-390$km/s at an absolute distance of about $\gtrsim8$~kpc, or for a fixed absolute distance of 3~kpc at a relative velocity of $\gtrsim460$~km/s, or an intermediate case). Thus, we cannot conclusively determine whether Hebe is bound to GN-z11 or not.

\section{Correspondence of  \texorpdfstring{H$\gamma$}{Hg} and He{\sc{ii}} spectra}\label{a:velmatch}

\begin{figure}
\centering
        \includegraphics[width=0.9\columnwidth]{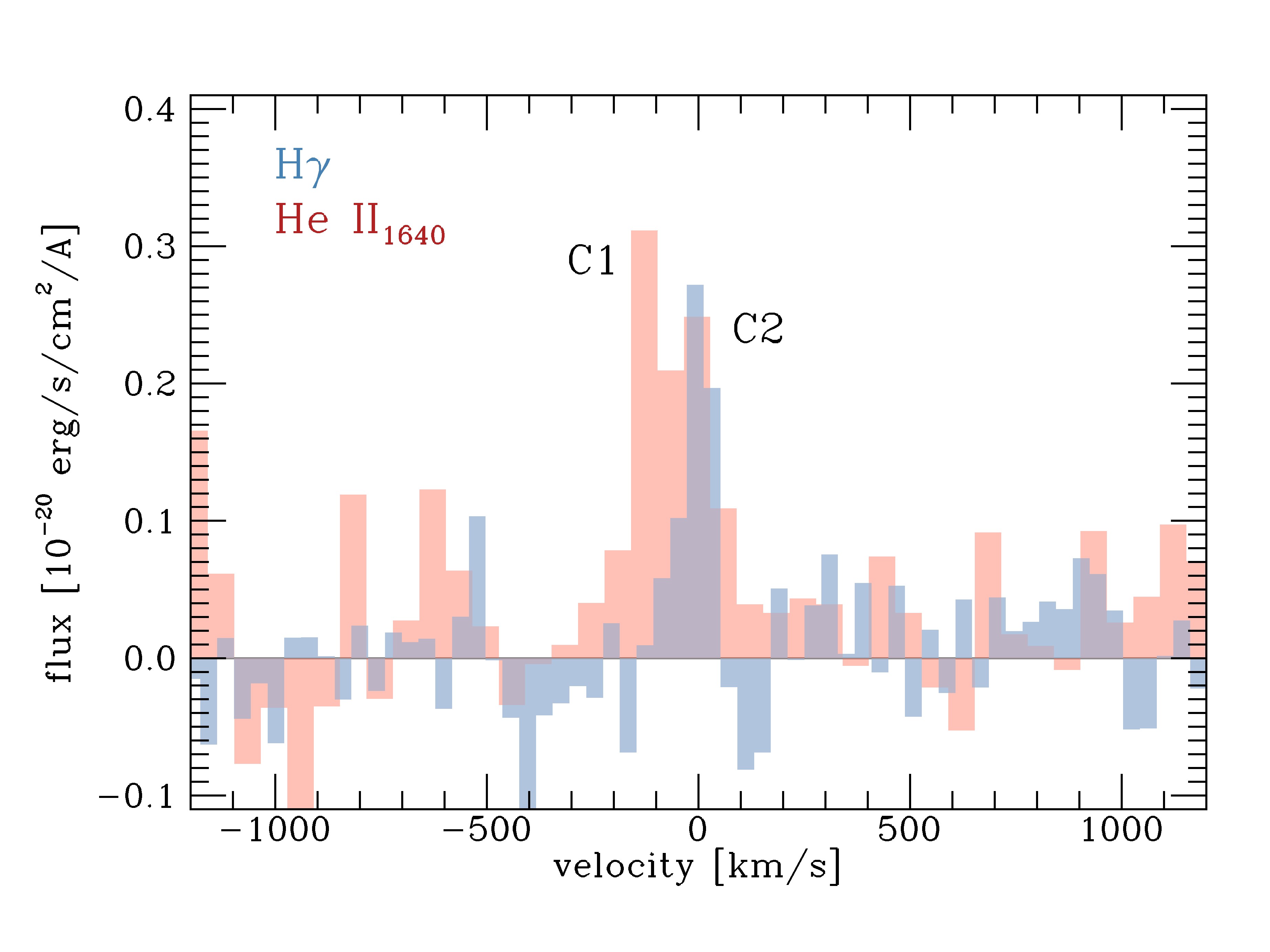}
    \caption{\small Overlay of the \hg\ and \heiiuv\ emission in velocity space (zero velocity is with reference to $z$=10.5862 for both lines). The \heiiuv\ spectrum is taken from \citetalias{Maiolino26b}.}
    \label{f:hgheiivel}
\end{figure}

In Fig.~\ref{f:hgheiivel} we show the \hg\ (blue) and \heiiuv\ (red; by \citetalias{Maiolino26b}) emission extracted from the $0.1^{\prime\prime}$$\times$$0.1^{\prime\prime}$ aperture shown in Fig.~\ref{f:map_extraction} in velocity space, where zero velocity corresponds to $z$=10.5862 for both lines. A second, bluer component is detected in \heiiuv\ (named C1 in the work by \citetalias{Maiolino26b} and \citealt{Rusta26}), while the red \heiiuv\ peak (C2) aligns with the \hg\ emission.

\section{Acknowledgements}

{\scriptsize We thank the anonymous referee for a constructive report that helped to improve the quality of this manuscript.
We thank Natascha~M.~F\"orster Schreiber for helpful discussions.
H\"U thanks the Max Planck Society for support through the Lise Meitner Excellence Program. H\"U acknowledges funding by the European Union (ERC APEX, 101164796). 
RM, GCJ, FDE and JS acknowledge support by the Science and Technology Facilities Council (STFC), by the ERC through Advanced Grant 695671 ``QUENCH'', and by the UKRI Frontier Research grant RISEandFALL. RM acknowledges funding from a research professorship from the Royal Society.
PGP-G acknowledges support from grant PID2022-139567NB-I00 funded by Spanish Ministerio de Ciencia, Innovaci\'on y Universidades MCIU/AEI/10.13039/501100011033,FEDER {\it Una manera de hacer Europa}.
YI is supported by JSPS KAKENHI Grant No.~24KJ0202.
MP acknowledges support through the grants PID2021-127718NB-I00, PID2024-159902NA-I00, and RYC2023-044853-I, funded by the Spain Ministry of Science and Innovation/State Agency of Research MCIN/AEI/10.13039/501100011033 and El Fondo Social Europeo Plus FSE+.
SA acknowledges support from grant PID2021-127718NB-I00 funded by Spanish Ministerio de Ciencia e Innovaci\'on MCIN/AEI/10.13039/501100011033
AJB and JC acknowledges funding from the ``FirstGalaxies'' Advanced Grant from the European Research Council (ERC) under the European Union's Horizon 2020 research and innovation programme (Grant agreement No.~789056).
SCa, GV and SZ acknowledge support by European Union's HE ERC Starting Grant No.~101040227 - WINGS.
EB and GC acknowledge funding from INAF ``Ricerca Fondamentale 2024'' (GO grant ``A JWST/MIRI MIRACLE: Mid-IR Activity of Circumnuclear Line Emission'' and mini-grant 1.05.24.07.01). 
ECL acknowledges support of an STFC Webb Fellowship (ST/W001438/1).
DJE is supported as a Simons Investigator.  DJE and BDJ were supported by JWST/NIRCam contract to the University of Arizona, NAS5-02105. 
BRP acknowledges support from grants PID2021-127718NB-I00 and PID2024-158856NA-I00 funded by Spanish Ministerio de Ciencia, Innovaci\'on y Universidades MCIU/AEI/10.13039/501100011033 and by ``ERDF A way of making Europe''.
ST acknowledges support by the Royal Society Research Grant G125142.
CNAW acknowledges a JWST/NIRCam contract to the University of Arizona (NAS5-02105).
JW gratefully acknowledges support from the Cosmic Dawn Center through the DAWN Fellowship. The Cosmic Dawn Center (DAWN) is funded by the Danish National Research Foundation under grant No.~140.
Views and opinions expressed are those of the authors only and do not necessarily reflect those of the European Union or the European Research Council Executive Agency. Neither the European Union nor the granting authority can be held responsible for them.
This work is based on observations made with the National Aeronautics and Space Administration (NASA)/European Space Agency (ESA)/Canadian Space Agency (CSA) {\it James Webb} Space Telescope. The data were obtained from the Mikulski Archive for Space Telescopes at the Space Telescope Science Institute, which is operated by the Association of Universities for Research in Astronomy, Inc., under NASA contract NAS 5-03127 for JWST. These observations are associated with program \#4528 (DOI:\href{https://doi.org/0.17909/hdrc-kv32}{10.17909/hdrc-kv32}).
}

\end{appendix}

\end{document}